\documentclass[12pt,twoside,a4paper]{article}
\usepackage{amsmath,amssymb,latexsym,theorem,bbm,shapepar,natbib,epsfig,color}
\newfont{\msa}{msam10 scaled\magstep1}
\newfont{\ssmsa}{msam9}

\def\DS{\mathop{\hbox{\rm DS}}}
\def\ES{\mathop{\hbox{\rm ES}}}
\def\EE{\mathop{\hbox{\rm EE}}}

\numberwithin{equation}{section}

\title{Bivariate ensemble model output statistics
  approach for joint forecasting of wind speed and temperature}

\author{{\sc S\'andor Baran$^{1}$} and {\sc Annette M\"oller$^{2}$}\\
         $^1$Faculty of Informatics, University of Debrecen\\
         Kassai \'ut 26, H-4028 Debrecen, Hungary \\
         $^2$ Department of Animal Sciences, University of
         G\"ottingen\\
         Carl-Sprengel-Weg 1, D-37075 G\"ottingen, Germany}

\date{}

\begin{document}
\pagestyle{myheadings}

\maketitle

\begin{abstract}
Forecast ensembles are typically employed to account for prediction
uncertainties in numerical weather prediction models. However,
ensembles often exhibit biases and dispersion errors, thus they
require statistical post-processing to improve their predictive
performance. Two popular univariate post-processing models are the
Bayesian model averaging (BMA) and the ensemble model output
statistics (EMOS).

In the last few years increased interest has emerged in developing multivariate
post-processing models, incorporating dependencies between weather
quantities, such as for example a bivariate distribution for wind
vectors or even a more general setting allowing to combine any types
of weather variables.

In line with a recently proposed approach to model temperature and
wind speed jointly by a bivariate BMA model, this paper introduces a
bivariate EMOS model for these weather quantities based
on a truncated normal distribution.

The bivariate EMOS model is applied to temperature and wind speed
forecasts of the eight-member University of Washington mesoscale
ensemble and of the eleven-member ALADIN-HUNEPS ensemble
of the Hungarian Meteorological Service and its predictive
performance is compared to the performance of the
bivariate BMA model and a multivariate Gaussian copula approach,
post-processing the margins with univariate EMOS. While the predictive
skills of the compared methods are similar, the bivariate EMOS model
requires considerably lower computation times than the bivariate BMA
method.

\bigskip
\noindent {\em Key words:\/} Ensemble model output statistics, Gaussian copula,
energy score, ensemble calibration, Euclidean error, truncated normal
distribution.
\end{abstract}

\section{Introduction}
  \label{sec:sec1}
Accurate and reliable prediction of future states of the
atmosphere is the most important objective of weather
prediction. These forecasts are issued on the basis of observational
data and numerical weather prediction (NWP) models, which
are capable to simulate the atmospheric motions taking
into account the physical governing laws of the atmosphere and the connected
spheres (typically sea or land surface). The NWP models consist of
sets of partial differential equations which have only numerical
solutions and strongly depend on initial conditions. In order to
reduce the uncertainties caused by the possibly unreliable initial
conditions and the numerical weather prediction process itself, one
can run the
models  with various initial conditions resulting in a forecast
ensemble \citep{leith}.  Using a forecast ensemble not only the
classical point forecasts (ensemble median or mean) can be obtained, but
also an estimate of the distribution of the future weather variable,
which allows probabilistic forecasting \citep{gr}. The first
operational implementation of
the ensemble prediction method dates back to the nineties
\citep{btmp,tk} and in the last twenty years it became a widely used
technique in the meteorological community. Recently all major national
meteorological services operate their own ensemble prediction systems
(EPSs), see, e.g., the PEARP\footnote{PEARP: Pr\'evision d'Ensemble ARPege}
EPS of M\'eteo France \citep{dljbac} or the
COSMO-DE\footnote{COSMO: Consortium for Small scale Modeling} EPS of
the German Meteorological Service \citep[DWD;][]{btg},
whereas the most well-known organization issuing ensemble forecasts is the
European Centre for Medium-Range Weather Forecasts
\citep{ecmwf}. However, as it has been observed with several operational
EPSs \citep[see, e.g.,][]{bhtp}, the forecast ensemble is usually
underdispersive and consequently badly calibrated. One possible
improvement area of the ensemble forecasts is the statistical
post-processing of the ensemble in order to transform the original
ensemble member-based probability density function (PDF) into a more
reliable and realistic one.

From the various post-processing techniques \citep[for an overview see,
e.g.,][]{gneiting14,wfk} probably the most popular approaches are the
Bayesian model averaging \citep[BMA;][]{rgbp} and the ensemble model
output statistics (EMOS) or non-homogeneous regression
\citep{grwg}. These methods are partially
implemented in the {\tt ensembleBMA} and {\tt ensembleMOS}
packages of {\tt R} \citep{frgsb} and both approaches provide estimates of the
distributions of the predictable weather quantities.

In the case of the BMA the predictive probability density function
(PDF) of a future weather quantity is a weighted mixture of individual
PDFs corresponding to the members of the ensemble, where the weights express
the relative performance of the ensemble members during a given
training period. The BMA models of various weather quantities differ only
in the PDFs of the mixture components. For temperature and sea level pressure
a normal distribution \citep{rgbp}, for wind speed a gamma
\citep{sgr10} or a truncated normal distribution \citep{bar}, whereas
for surface wind direction a von Mises distribution \citep{bgrgg} is
suggested.

The EMOS predictive PDF uses a single parametric distribution with
parameters depending on the ensemble members. EMOS models have already
been developed for calibrating ensemble forecasts of temperature and
sea level pressure \citep{grwg}, wind speed \citep{tg,lt,bl} and
precipitation \citep{sch}.

Besides the calibration of univariate weather quantities recently an
increasing interest has appeared in modeling correlations between the
different weather variables. In the special case of wind vectors
\citet{pinson} suggested an adaptive calibration technique, whereas
\citet{stg} and \citet{sgr13} introduced bivariate EMOS and BMA
models, respectively. Further, \citet{mlt} developed a general approach
where after univariate calibration of the weather variables
the component predictive PDFs are joined
into a multivariate predictive density with the help of a Gaussian
copula. Another idea appears in the ensemble copula coupling (ECC)
method \citet{stg13} where after univariate calibration the rank order
information in the raw ensemble is used to restore
correlations. Finally, \citet{bm} developed a BMA model for joint
post-processing of ensemble forecasts of wind speed and temperature.

In the present paper we introduce an EMOS model for joint calibration
of wind speed and temperature which is based on a truncated normal distribution
with cut-off at zero in its first (wind) coordinate. The method is
tested on the ensemble forecasts of wind speed and temperature of the
eight-member University of Washington
Mesoscale Ensemble \citep[UWME;][]{em05} and of the Limited Area Model
EPS of the Hungarian
Meteorological Service (HMS) called ALADIN-HUNEPS\footnote{ALADIN:
  Aire Limit\'ee Adaptation dynamique Development International}
\citep{horanyi}. The
performance of the EMOS model is compared to the forecasting skills of
the previously investigated BMA method of \citet{bm} and to the
Gaussian copula approach of \citet{mlt}, where the margins of the multivariate
predictive distribution are estimated by EMOS.

\section{Data}
  \label{sec:sec2}

\subsection{University of Washington mesoscale ensemble}
  \label{subs:subs2.1}
The eight-member University of Washington mesoscale ensemble covers the Pacific
Northwest region of western North America providing forecasts on a 12
km grid. The ensemble members are
obtained from different runs of the fifth generation Pennsylvania State
University--National Center for Atmospheric Research mesoscale model
(PSU-NCAR MM5) with initial conditions from different sources
\citep{grell}. Our data base (identical to the one used in
\citet{mlt,bm}) contains ensembles of 48-hour forecasts
and corresponding validation observations of 10 meter maximum wind
speed (maximum of the hourly instantaneous wind speeds over the
previous twelve hours, given in m/s, see, e.g., \citet{sgr10}) and 2
meter minimum temperature (given in K) for 152 stations in the Automated Surface
Observing Network \citep{asos} in the US states of Washington, Oregon,
Idaho, California and Nevada for calendar years 2007 and 2008. The
forecasts are
initialized at 0 UTC (5 pm local time when daylight saving time (DST) is in use
and 4 pm otherwise) and the generation of the ensemble implies that
its members are not exchangeable. In the present study we investigate
only forecasts for calendar year 2008 with additional data from  2007
used for parameter estimation. After removing
days and locations with missing data, 90 stations remained where
the number of days for which forecasts and validating observations are
available varies between 141 and 290.

\subsection{ALADIN-HUNEPS ensemble}
  \label{subs:subs2.2}
The ALADIN-HUNEPS system of the HMS covers a large part of Continental
Europe with a
horizontal resolution
of 8 km and it is obtained by dynamical downscaling (by the ALADIN
limited area model) of the global
ARPEGE\footnote{ARPEGE: Action de Recherche Petite Echelle Grande
  Echelle} based PEARP system of M\'et\'eo France \citep{hkkr,dljbac}. The
ensemble consists of 11 members, 10 initialized from perturbed initial
conditions and one control member from the unperturbed analysis,
implying that the ensemble contains groups of exchangeable
forecasts. The data base contains 11 member ensembles of 42-hour
forecasts for 10 meter instantaneous wind speed (given
in m/s) and 2 meter temperature (given in K) for 10 major cities in
Hungary (Miskolc, Szombathely, Gy\H or, Budapest, Debrecen, Ny\'\i regyh\'aza,
Nagykanizsa, P\'ecs, Kecskem\'et, Szeged) produced by the
ALADIN-HUNEPS system of the HMS, together with the corresponding
validating observations for the one-year period between April 1, 2012
and March 31, 2013 and for the period from October 1, 2010 to March
25, 2011. The forecasts are initialized at 18 UTC (8 pm local time when DST
operates and 7 pm otherwise). The data sets are fairly complete since there are
only six and three days, respectively, when no forecasts are available
and these days have been excluded from the analysis.

\section{Ensemble Model Output Statistics}
  \label{sec:sec3}

As mentioned in the Introduction, the EMOS predictive PDF of a
weather quantity (vector) \ $X$ \ is a single parametric density function
where the parameters depend on the ensemble members. For temperature
and pressure a normal distribution can be fit reasonably well
\citep{grwg}, while for wind vectors a bivariate normal distribution
can be applied \citep{stg}. However, for modeling non-negative quantities such
as wind speed, a skewed distribution is required. \citet{tg}
introduced an EMOS model based on truncated normal distribution with
cut-off at zero, but EMOS models utilizing a generalized extreme value
distribution \citep{lt} and a log-normal distribution \citep{bl} have
also been tested.  The EMOS models of \citet{grwg} and \citet{tg}
suggest the idea of joint modeling wind speed and temperature using a
bivariate normal distribution with first (wind) coordinate truncated
from below at zero. This particular distribution has already been
applied in the bivariate BMA model of \citet{bm}.

Once the predictive density is given, its mean or median can be taken
as a point forecast for \ $X$. \ In one dimension the definition of
the latter is obvious, whereas for a $d$-dimensional cumulative distribution
function (CDF) \ $F$ \ a
multivariate median is a vector minimizing the function
\begin{equation*}
\phi(\boldsymbol\alpha):=\int_{\mathbb R^d} \Vert \boldsymbol\alpha -\boldsymbol
x\Vert F({\mathrm d}\boldsymbol x),
\end{equation*}
where \ $\Vert\cdot\Vert$ \ denotes the Euclidean norm. If \ $F$ \ is
not concentrated on a line in \ $\mathbb R^d$ \ then the median is
unique \citep{md}.

Denote by \ $\boldsymbol f_1,\boldsymbol f_2,\ldots , \boldsymbol f_M$
\ the ensemble of
distinguishable forecast vectors of wind speed and temperature for a
given location and time. This means that each ensemble member can be
identified and tracked, which holds for example for the UWME (see Section
\ref{subs:subs2.1}).
However, most of the currently used ensemble prediction systems
provide ensembles where at least some members are statistically
indistinguishable. Such ensemble systems are simulating
uncertainties by perturbing the initial conditions, and they usually
have a control member (the one without any perturbation), whereas
the remaining ensemble members form one or two exchangeable
groups.  This is the case, e.g., for the 51 member
ECMWF ensemble \citep{lp} or for the
ALADIN-HUNEPS ensemble described in Section \ref{subs:subs2.2}.

In what follows, if we have \ $M$ \ ensemble members divided
into \ $m$ \ exchangeable
groups, where the \ $k$th \ group contains \ $M_k\geq 1$ \ ensemble
members ($\sum_{k=1}^mM_k=M$), \ notation \ $\boldsymbol f_{k,\ell}$
\ will be used for the  $\ell$th member of the $k$th group.

\subsection{Bivariate truncated normal model}
  \label{subs:subs3.1}

Denote by \ ${\mathcal N}_2^{\, 0}(\boldsymbol \mu, \Sigma)$ \ the bivariate
normal distribution with location vector \ $\boldsymbol \mu$, \ scale
matrix \ $\Sigma$, \ and first coordinate truncated from below at zero.
Let
\begin{equation*}
\boldsymbol\mu=\begin{bmatrix}\mu_W \\ \mu_T \end{bmatrix} \qquad
\text{and} \qquad \Sigma=\begin{bmatrix}\sigma^2_W& \sigma_{WT}
  \\\sigma_{WT}&\sigma^2_{T} \end{bmatrix}.
\end{equation*}
If \  $\Sigma$ \ is regular, then the PDF of this distribution is
\begin{equation}
  \label{eq:eq3.1}
g(\boldsymbol x| \boldsymbol\mu,\Sigma )\!:=\!\frac
{\big(\det(\Sigma)\big)^{-1/2}}
{2\pi\Phi\big(\mu_W/\sigma_W\big)}\exp\Big(-\frac
12(\boldsymbol x-\boldsymbol\mu)^{\top}\Sigma^{-1}(\boldsymbol
x-\boldsymbol\mu)\Big){\mathbb I}_{\{x_W\geq 0\}}, \quad \boldsymbol
x\!=\begin{bmatrix}x_W \\x_T \end{bmatrix}\! \in\! {\mathbb R}^2,
\end{equation}
where \ $\Phi$ \ denotes the CDF of the
standard normal distribution and by \ ${\mathbb I}_H$ \ we denote the indicator
function of a set \ $H$. \ Short calculation shows \citep[see,
e.g.,][]{rose}, that the mean vector \ $\boldsymbol\kappa$ \ and
covariance matrix \ $\Xi$ \ of \
\ ${\mathcal N}_2^{\, 0}(\boldsymbol \mu, \Sigma)$  \ are
\begin{align*}
\boldsymbol\kappa&=\boldsymbol\mu
+\frac{\varphi\big(\mu_W/\sigma_W\big)}{\Phi\big(\mu_W/\sigma_W\big)}
\begin{bmatrix}\sigma_W \\\sigma_{WT}/\sigma_W \end{bmatrix}
 \qquad \qquad \text{and} \\
\Xi&=\Sigma-\left(
  \frac{\mu_W}{\sigma_W}\frac{\varphi\big(\mu_W/\sigma_W\big)}{\Phi
    \big(\mu_W/\sigma_W\big)}
  +\Bigg(\frac{\varphi\big(\mu_W/\sigma_W\big)}{\Phi
    \big(\mu_W/\sigma_W\big)}\Bigg)^2\right) \begin{bmatrix}\sigma^2_W&
  \sigma_{WT}
  \\\sigma_{WT}& \sigma^2_{WT}/\sigma^2_W \end{bmatrix} ,
\end{align*}
respectively, where \ $\varphi$ \ denotes the PDF of the
standard normal distribution.

The proposed EMOS predictive distribution of wind speed and
temperature is
\begin{equation}
   \label{eq:eq3.2}
  {\mathcal N}_2^{\, 0}\big(A+B_1\boldsymbol f_1+ \cdots
  +B_M \boldsymbol f_M,C+DS D^{\top}\big)
\end{equation}
with
\begin{equation*}
  S:=\frac 1{M-1}\sum_{k=1}^M\big (\boldsymbol
  f_k-\overline{\boldsymbol f}\big)\big (\boldsymbol f_k-\overline
  {\boldsymbol f}\big)^{\top},
\end{equation*}
where \ $\overline{\boldsymbol f}$ \ denotes the ensemble mean
vector. Parameter vector
\ $A\in {\mathbb R}^2$ \ and two-by-two real parameter matrices \
$B_1, \ldots , B_M$ \ and \ $C,\ D $ \ of model \eqref{eq:eq3.2}, \
where $C$ is assumed to be symmetric and non-negative definite, can be
estimated from the training data consisting of ensemble members and verifying
observations from the preceding \ $n$ \ days, by optimizing with
respect to the mean logarithmic score, i.e., the negative logarithm of
the predictive PDF evaluated at the verifying observation \citep{gsghj}.
We remark that under the assumption of independence in space and time,
this approach is equivalent to the maximum likelihood
method. Obviously, the forecast errors are usually not independent,
however, since one is estimating the conditional
distribution of a single weather quantity vector with respect to the
corresponding forecasts, the parameter estimates are not really
sensitive to this assumption \citep[see, e.g.,][]{rgbp}.

If the ensemble can be divided into groups of
exchangeable members, ensemble members within a given group will get the
same coefficient matrix of the location parameter \citep{frg,gneiting14}
resulting in a predictive distribution of the form
\begin{equation}
   \label{eq:eq3.3}
  {\mathcal N}_2^{\, 0}\Bigg(A+B_1\sum_{\ell_1=1}^{M_1}\boldsymbol
  f_{1,\ell_1}+ \cdots  +B_m \sum_{\ell_m=1}^{M_m}\boldsymbol
  f_{m,\ell_m},C+DS D^{\top}\Bigg),
\end{equation}
where  again, \ $S$ \ denotes the empirical covariance matrix of the
ensemble.

\subsection{Verification scores}
 \label{subs:subs3.2}

To investigate the predictive skills of the probabilistic and point
forecasts we apply the multivariate scores proposed by \citet{gsghj}.

The first step is to check the calibration of probabilistic
forecasts, which notion means a statistical consistency between the predictive
distributions and the observations \citep[see, e.g.,][]{tg}.
For one-dimensional ensemble forecasts a frequently used tool for
this purpose is the verification rank histogram, i.e., the
histogram of ranks of validating observations with respect to the
ensemble forecasts \citep[see, e.g.,][Section 8.7.2]{wilks}. The
closer the distribution of the ranks to the uniform distribution on \
$\{1,2, \ldots ,M+1\}$, \ the better the calibration. The deviation
from uniformity can be quantified by the reliability index \ $\Delta$
\ defined as
\begin{equation*}
 \Delta:=\sum_{r=1}^{M+1}\Big| \rho_r-\frac 1{M+1}\Big|,
\end{equation*}
where \ $\rho_j$ \ is the relative frequency of rank \ $r$ \
\citep{delle}. In the multivariate case the proper definition of ranks
is not obvious. Similar to \citet{bm}, in the present work we use the
multivariate ordering proposed by \citet{gsghj}. For a probabilistic
forecast one can calculate the reliability index from a preferably
large number of ensembles (we use $100$) sampled from the predictive
PDF and the corresponding verifying observations.

For evaluating multivariate density forecasts the most popular scoring rules
are the logarithmic score and the energy score (ES), introduced by
\citet{grjasa}.  Both the logarithmic and the energy score are proper
scoring rules which are negatively oriented, that
is, the smaller the better, and the latter is a direct multivariate
extension of the continuous ranked probability score (CRPS). Given a
predictive CDF \ $F$ \ on \
${\mathbb R}^d$ \ and a $d$-dimensional observation \ $\boldsymbol x$, \
the energy score is defined as
\begin{equation*}
\ES(F,\boldsymbol x):={\mathsf E}\Vert \boldsymbol X-\boldsymbol
x\Vert-\frac 12 {\mathsf E}\Vert \boldsymbol X-\boldsymbol X'\Vert,
 \end{equation*}
where \ $ \boldsymbol X$ \ and \  $\boldsymbol X'$ \ are independent
random vectors with CDF \ $F$. \ However, for the
bivariate truncated normal distribution the
energy score cannot be given in a closed form, so it is replaced by a
Monte Carlo approximation
\begin{equation}
  \label{eq:eq3.4}
\widehat\ES(F,\boldsymbol x):=\frac 1n \sum _{j=1}^n\Vert \boldsymbol
X_j-\boldsymbol x\Vert-\frac 1{2(n-1)}\sum_{j=1}^{n-1}\Vert
\boldsymbol X_j-\boldsymbol X_{j+1}\Vert,
 \end{equation}
where \ $\boldsymbol X_1,\boldsymbol X_2, \ldots ,\boldsymbol X_n$ \
is a (large, we use \ $n=10000$) \ random sample from \ $F$ \
\citep{gsghj}. Finally, if \
$F$ \ is a CDF corresponding to a forecast ensemble \ $\boldsymbol
f_1,\boldsymbol f_2, \ldots ,\boldsymbol f_M$ \ then \eqref{eq:eq3.4}
reduces to
\begin{equation*}
\ES(F,\boldsymbol x)=\frac 1M \sum _{j=1}^M\Vert \boldsymbol
f_j-\boldsymbol x\Vert-\frac 1{2M^2}\sum_{j=1}^M\sum_{k=1}^M\Vert
\boldsymbol f_j-\boldsymbol f_k\Vert.
 \end{equation*}

Besides the proper calibration, probabilistic forecasts should result
in sharp predictive distributions. In the univariate case this usually
means small standard deviations leading to narrow central prediction
intervals. For a $d$-dimensional quantity one can consider the
determinant sharpness \ ($\DS$) \ defined by
\begin{equation*}
\DS:= \big(\det (\Sigma)\big)^{1/(2d)},
\end{equation*}
where \ $\Sigma $ \ is the covariance matrix of an ensemble or of a
predictive PDF.

Finally, point forecasts (median and mean) can be evaluated using
the mean Euclidean distance \ ($\EE$) \ of forecasts from the corresponding
validating observations. For multivariate forecasts the ensemble
median can be obtained, e.g.,  using the Newton-type algorithm given in
\citet{ds83} or the algorithm of \citet{vz}. For a detailed comparison
of different algorithms, see, e.g., \citet{ffc}.
Given a predictive CDF,  to determine the corresponding median the chosen
algorithm might be applied on a preferably large sample from this distribution.

\subsection{Parameter estimation}
  \label{subs:subs3.3}

There are two possible approaches to the choice of training data for
estimating the unknown parameters of the various EMOS models
\citep{tg,stg}. The regional EMOS technique uses ensemble forecasts
and validating observations from a rolling training period for all available
stations. In this way, one gets a universal set of
parameters across the entire ensemble domain, which is then used at all
observation sites. E.g., in case of the ALADIN-HUNEPS ensemble this
means a single set of parameters for all ten cities. In contrast,
local EMOS produces distinct parameter estimates for the different
stations by using only the training data of the given station. These
training sets contain only one observation per day, so local EMOS
models require long training periods.

Now, e.g., in the bivariate model \eqref{eq:eq3.3} the number of free
parameters to be estimated is $4m+10$, which means 14 parameters even
in the simplest case of a single exchangeable ensemble group. Hence,
for estimating the parameters of models \eqref{eq:eq3.2}
and \eqref{eq:eq3.3} only the regional EMOS approach is applicable.

The mean logarithmic score is optimized numerically with the help of the
{\tt optim} function in {\tt R}, using principally the Nelder-Mead
\citep{nm} algorithm. This method is slower but more robust than the
popular Broyden-Fletcher-Goldfarb-Shanno (BFGS) algorithm
\citep[Section 10.9]{press}, which in case of a small training set
becomes unstable. Both optimization methods require initial values, and
the starting values of the location parameters \ $A$ \ and \ $B_1,\ldots
,B_M$ \ are coefficients of the bivariate linear regression of the
observations on the ensemble forecasts over the training period.
Further, for the scale parameters \ $C$ \ and \ $D$, \ the previous
day's estimates can serve as initials values, however, according to
our experience, fixed starting values provide slightly better results. Finally,
to enforce the non-negative definiteness of the parameter matrix \ $C$
\ one can set \ $C={\mathcal C}{\mathcal C}^{\top}$ \ and perform the
optimization with respect to \ ${\mathcal C}$.

\section{Results}
  \label{sec:sec4}

As mentioned in the Introduction, the predictive performance
of the bivariate EMOS model (see Section \ref{subs:subs3.1}) is tested
on the eight-member UWME and on the ALADIN-HUNEPS ensemble of the
HMS. The goodness of fit of the predictive distributions is quantified
with the multivariate scores given in Section \ref{subs:subs3.2}, and
the obtained results are compared to the fits of the independent EMOS
models of wind speed \citep{tg} and temperature \citep{grwg}, the
Gaussian copula method proposed by \citet{mlt}, but with marginal
distributions estimated by EMOS models, and the bivariate BMA
model of \citet{bm}. We remark that the parameters of the independent
univariate EMOS models are estimated by minimizing the mean CRPS of
the training data. For fitting the marginal predictive distributions
in the Gaussian copula approach, we employ the same univariate EMOS
models for wind speed and temperature as in the independent
approach. Therefore, their model parameters are estimated by the
minimum CRPS method as well.
If one has a closed expression for the CRPS, which
is the case both for the normal and the truncated normal distribution,
this method usually gives better results than optimization with
respect to the logarithmic score.

\subsection{University of Washington Mesoscale Ensemble}
\label{subs:subs4.1}

\subsubsection{Raw ensemble}
 \label{subsub4.1.1}

\begin{figure}[t]
\begin{center}
\leavevmode
\epsfig{file=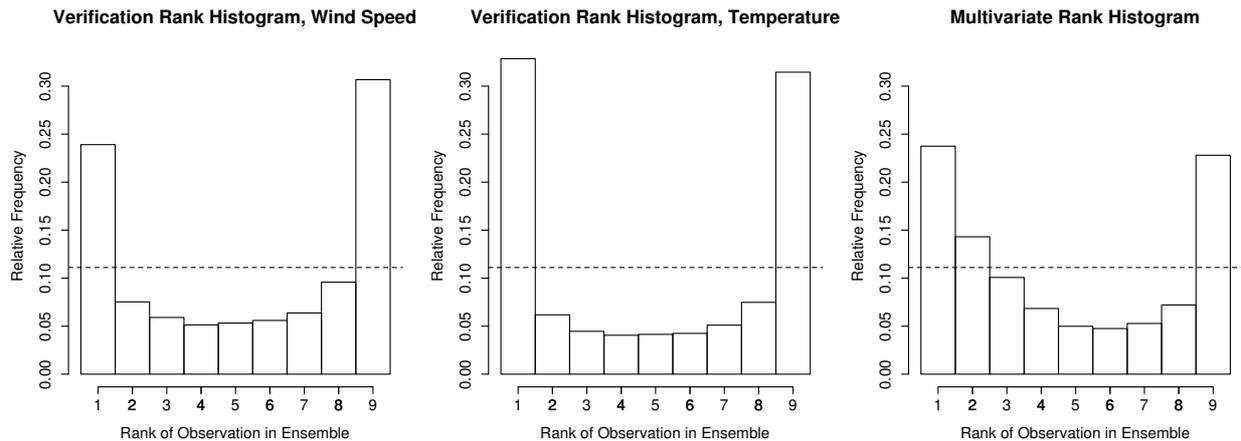,height=5.7cm, angle=0}
\caption{Verification rank histograms of the 8-member UMWE forecasts
  of maximum wind speed ({\em left}) and minimum temperature ({\em
    center}) and the multivariate rank histogram ({\em right}).
  Period: January 1,  2008 --  December 31, 2008.}
\label{fig:fig1}
\end{center}
\end{figure}
Several studies have verified that wind speed and temperature
forecasts of the UWME are strongly underdispersive \citep[see,
e.g.,][]{tg,frg}, and consequently uncalibrated. Obviously, the lack
of calibration will remain valid if one considers these ensemble
forecasts together, as predictions of a bivariate weather quantity
\citep{bm}. The underdispersive character of the raw ensemble can
nicely be observed in Figure \ref{fig:fig1} (identical to Figure 1 of
\citet{bm}) displaying the univariate verification rank
histograms of wind speed and temperature forecasts together with their
joint multivariate rank histogram. The corresponding reliability
indices \ $\Delta$ \ are $0.647, \ 0.842$ and $0.550$, respectively, and
in many cases the raw ensemble either over-, or underestimates the
verifying observation. Further, the need of bivariate modeling can be
justified both by the positive correlation of $0.125$ of the verifying
observations of wind speed and temperature for calendar year 2008
taken along all dates and locations, and by the correlations of
$0.187$ and $0.189$ of forecast errors of the ensemble median and
mean, respectively.

\subsubsection{Bivariate EMOS calibration}
 \label{subsub4.1.2}
\begin{table}[t!]
\begin{center}{
\begin{tabular}{lccccccccc}
&\multicolumn{3}{c}{Probabilistic forecasts}&\multicolumn{3}{c}{Median
  forecasts}&\multicolumn{3}{c}{Mean forecasts} \\ \hline
&$\ES$&$\Delta$&$\DS$&$\EE$&$\varrho$&$\varrho_{err}$&$\EE$&$\varrho$&$
\varrho_{err}$\\ \hline
EMOS&$2.127$&$0.025$&$2.273$&$2.982$&$0.165$&$0.182$&$2.982$&$0.157$&$
0.182$\\ 
Indep. EMOS&$2.118$&$0.059$&$2.206$&$2.966$&$0.164$&$0.176$&$2.966$&$0.155
  $&$0.178$\\
\hline
Copula&$2.088$&$0.021$&$2.169$&$2.967$&$0.162$&$0.178$&$2.967$&$0.156$&$0.179$\\
BMA&$2.110$&$0.015$&$2.250$&$2.973$&$0.154$&$0.182$&$2.972$&$0.155$&$ 0.183$\\
\hline
Raw
ensemble&$2.562$&$0.550$&$0.773$&$3.087$&$0.017$&$0.187$&$3.072$&$0.007$&$
0.189$
\end{tabular}
\caption{Mean energy
  score ($\ES$), reliability index
  ($\Delta$) and mean determinant sharpness ($\DS$) of probabilistic
  forecasts, mean Euclidean  error ($\EE$) of point forecasts
  (median/mean), empirical correlation ($\varrho$) and empirical
  correlation of errors ($\varrho_{err}$) of wind speed and
  temperature  components of point forecasts for the
  UWME.  Empirical correlation of observations corresponding to the
  forecast cases: $0.125$.} \label{tab:tab1} }
\end{center}
\end{table}
The first step of EMOS (and BMA) post-processing of ensemble
forecast is the selection of the length of the rolling training period.
In order to ensure comparability of the results with the findings of
earlier studies, we apply the same 40 days training period length  as
in \citet{mlt} and \citet{bm}. This training period length was a
result of an exploratory data analysis on a subset of the data set.
Similar to the previous studies, we produce EMOS predictive PDFs for
the whole calendar year 2008, using also the data from the last two months
of calendar year 2007. After removing dates with missing data this means
291 calendar days with a total of 24\,302 forecast cases.
As the eight ensemble members of the UWME are not exchangeable, for
calibration we apply bivariate EMOS model \eqref{eq:eq3.2} with \ $M=8$.

In case of the copula method, the data from calendar year 2007 are
applied for estimating the correlation between the two weather
quantities, and the resulting correlation matrix is then carried
forward into the analysis of the 2008 data. This is in accordance with the
BMA based copula calibration of \citet{bm}.

\begin{figure}[t]
\begin{center}
\leavevmode
\epsfig{file=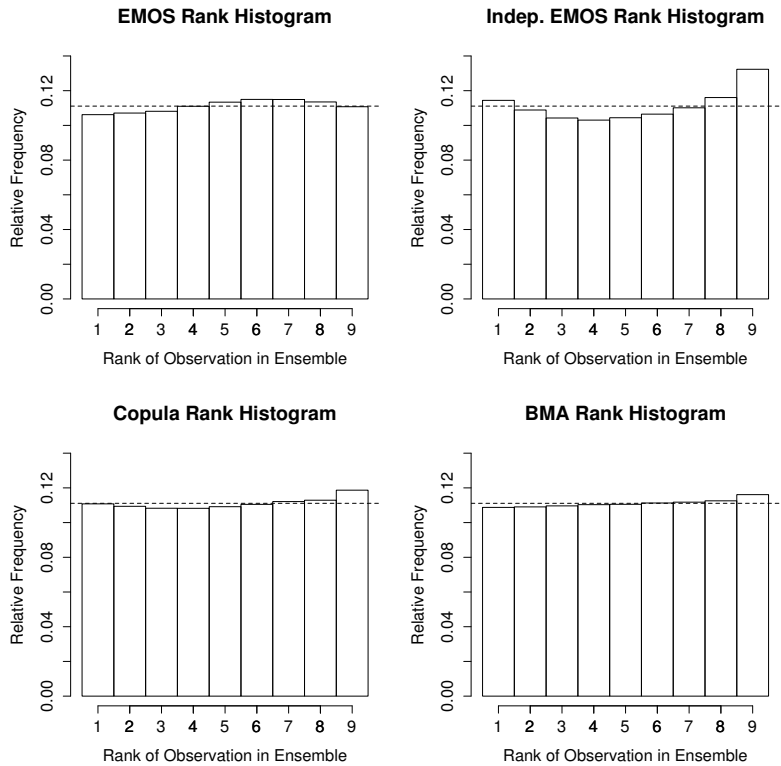,height=10cm, angle=0}
\caption{Multivariate rank histograms for EMOS ({\em upper left}),
  independent EMOS ({\em upper right}), Gaussian copula ({\em lower
    left}) and BMA ({\em lower right})
  post-processed UWME forecasts of maximum wind speed and minimum temperature.}
\label{fig:fig2}
\end{center}
\end{figure}

In Table \ref{tab:tab1} the verification scores calculated using the
EMOS model \eqref{eq:eq3.2}, the independent EMOS models of wind speed and
temperature, the copula model of \citet{mlt} with EMOS post-processed
margins, the BMA model of
\citet{bm} and the raw ensemble are given.  Compared to the raw
ensemble all post-processing techniques substantially improve the
calibration of probabilistic forecasts which
is quantified by the significant decrease of the mean energy score
($\ES$) and reliability index ($\Delta$) and can also be observed in
Figure \ref{fig:fig2} showing the rank histograms of post-processed
forecasts. These almost uniform histograms should be compared to the
rank histograms of the raw ensemble plotted in Figure
\ref{fig:fig1}. The price to pay for the better calibration is the loss in
sharpness (see the corresponding values of $\DS$), however, this is a
direct consequence of the small dispersion of the raw ensemble (see
again Figure \ref{fig:fig1}).
Post-processing also results in slightly
smaller mean Euclidean errors ($\EE$) indicating more accurate median
and mean forecasts. Further, the empirical correlations $\varrho$ of
the wind and temperature components of the post-processed point
forecasts are much closer to the correlation of $0.125$ of the
verifying observations than the corresponding correlations of the
ensemble median and mean. This latter is a weakness of the raw
ensemble, however, one should also remark that all error correlations
$\varrho_{err}$ (including the raw ensemble) are very similar to each
other (around $0.180$).

Comparing the different post-processing techniques one can observe
that the main difference between the various approaches appears in the
reliability index. The bivariate BMA model results in the smallest \
$\Delta$ \ value, followed by the copula and the bivariate EMOS methods,
which is in line with shapes of the multivariate rank histograms
plotted in Figure \ref{fig:fig2}. Further, the large \ $\Delta$ \
value and the U-shaped rank
histogram of the independent EMOS approach supports the idea of
bivariate modeling. However, in the model choice one should also take
into account that the copula method requires additional data for
estimating the correlation matrix, whereas in the BMA and EMOS approaches
the parameters are estimated using only the training data. Finally, in
case of the latter two
methods the computational costs (see Section \ref{subs:subs4.3}) might
also have an influence on the decision.

\subsection{ALADIN-HUNEPS Ensemble}
\label{subs:subs4.2}

\subsubsection{Raw ensemble}
 \label{subsub4.2.1}

\begin{figure}[t]
\begin{center}
\leavevmode
\epsfig{file=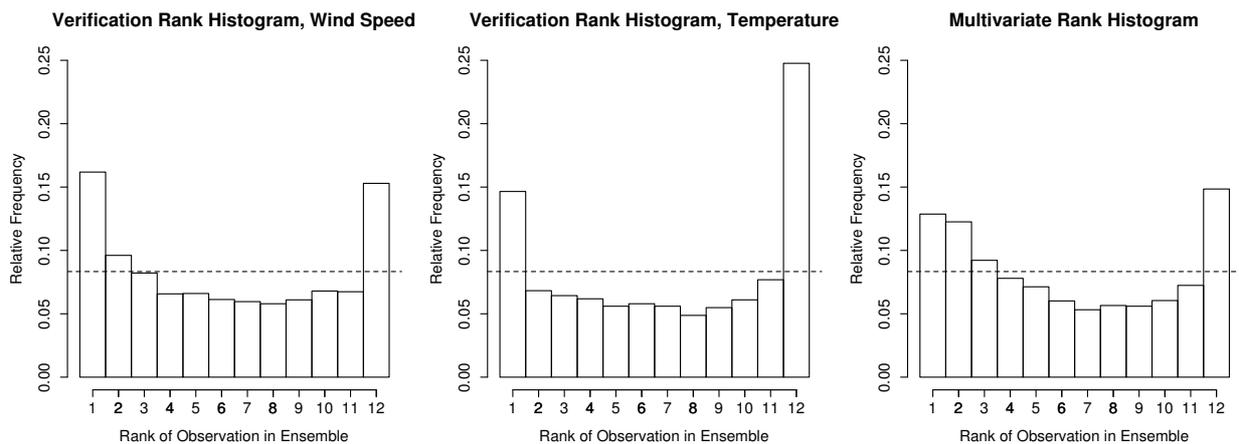,height=5.7cm, angle=0}
\caption{Verification rank histograms of the 11-member ALADIN-HUNEPS
  ensemble forecasts of  wind speed ({\em left}) and temperature ({\em
    center}) and the multivariate rank histogram ({\em right}).
  Period: April 1,  2012 --  March 31, 2013.}
\label{fig:fig3}
\end{center}
\end{figure}

Wind speed and temperature forecasts of the ALADIN-HUNEPS EPS are
better calibrated than those of the UWME, however, the rank histograms
in Figure \ref{fig:fig3} still exhibit a strong underdispersive
character. The bivariate reliability index equals $0.317$, whereas the
reliability indices of wind speed and temperature are $0.322$ and
$0.455$, respectively. The need of bivariate post-processing is again
supported by the forecast error correlations of $0.119$ and $0.123$ of the
ensemble median and mean, respectively, however, in this case the
verifying observations of wind speed and temperature show a very
slight negative correlation of $-0.029$. This latter difference
compared to the UWME, where this correlation equals $0.125$, might be
explained by the different types of wind and temperature quantities
being examined (see Sections \ref{subs:subs2.1} and \ref{subs:subs2.2}).

\subsubsection{Bivariate EMOS calibration}
 \label{subsub4.2.2}

Similar to the case of the UWME, to ensure the
comparability of the results with the bivariate BMA post-processing of the
same forecast data, we keep the 40-day training period of
\citet{bm}. This particular training period length was the outcome of
a preliminary data analysis consisting of univariate BMA and EMOS
calibration of wind speed and temperature forecasts. Hence,
ensemble forecasts, validating observations and predictive
distributions are available for the period from May 12, 2012 to March 31,
2013, which means 318 days and 3\,180 forecast cases as 6 days with missing
forecasts are excluded from the analysis.

\begin{table}[t!]
\begin{center}{
\begin{tabular}{lccccccccc}
&\multicolumn{3}{c}{Probabilistic forecasts}&\multicolumn{3}{c}{Median
  forecasts}&\multicolumn{3}{c}{Mean forecasts} \\ \hline
&$\ES$&$\Delta$&$\DS$&$\EE$&$\varrho$&$\varrho_{err}$&$\EE$&$\varrho$&$
\varrho_{err}$\\ \hline
EMOS
&$1.442$&$0.034$&$1.478$&$2.015$&$-0.041$&$0.132$&$2.016$&$-0.049$&$
0.132$\\
Indep. EMOS&$1.436$&$0.051$&$1.456$&$2.002$&$-0.033$&$0.128$&$2.002$&$
-0.044$&$0.127 $\\ \hline
Copula&$1.384$&$0.075$&$1.557$&$2.000$&$-0.036$&$0.128$&$2.000$&$-0.039$&$0.127
$\\
BMA&$1.434$&$0.031$&$1.539$&$2.004$&$-0.032$&$0.129$&$2.007$&$
-0.041$&$0.129$\\
\hline
Raw ensemble&$1.623$&$0.327$&$0.935$&$2.102$&$-0.068$&$0.122$&$2.083$&$
-0.060$&$0.124$
\end{tabular} }
\caption{Mean energy score ($\ES$), reliability index
  ($\Delta$) and mean determinant sharpness ($\DS$) of probabilistic
  forecasts, mean Euclidean  error ($\EE$) of point forecasts
  (median/mean), empirical correlation ($\varrho$) and empirical
  correlation of errors ($\varrho_{err}$) of wind speed and
  temperature  components of point forecasts for the  ALADIN-HUNEPS
  ensemble. Empirical correlation of observations  corresponding to the
  forecast cases: $-0.033$.} \label{tab:tab2}
\end{center}
\end{table}

Further, the way the ALADIN-HUNEPS ensemble is generated (see Section
\ref{subs:subs2.2}) induces a natural grouping of the ensemble members
into two groups. The first group contains just the control member \
$\boldsymbol f_c$, \ whereas in the second are the 10
statistically indistinguishable ensemble members \ $\boldsymbol
f_{p,1}, \ldots , \boldsymbol f_{p,10}$,\ initialized from
randomly perturbed initial conditions. This results in the predictive
PDF
\begin{equation}
  \label{eq:eq4.1}
{\mathcal N}_2^{\, 0}\Bigg(A+B_c \boldsymbol
  f_c+B_p \sum_{\ell=1}^{10}\boldsymbol
  f_{p,\ell},C+DS D^{\top}\Bigg),
\end{equation}
which is a special case of model \eqref{eq:eq3.3}. One should remark here that
in \citet{bhn1} a
different grouping is also suggested (and later investigated in
\citet{bar, bhn2} and \citet{bm}, too), where the odd and even numbered
exchangeable ensemble members form two separate groups. This idea is
justified by the method their initial conditions
are generated, since only five perturbations are calculated and
then they are added to (odd numbered members) and
subtracted from (even numbered members) the unperturbed
initial conditions. However, since in the present study the results
corresponding to the two- and three-group models are rather similar, only
the two-group case is reported.

In line with the similar case study of \citet{bm},
to estimate the correlation matrix of the Gaussian copula, additional
data of the period from October 1, 2010 to March 25, 2011 are
utilized, and the estimated correlation matrix is employed for combining the
univariate EMOS marginals for 2012/2013 in the Gaussian copula.

\begin{figure}[t]
\begin{center}
\leavevmode
\epsfig{file=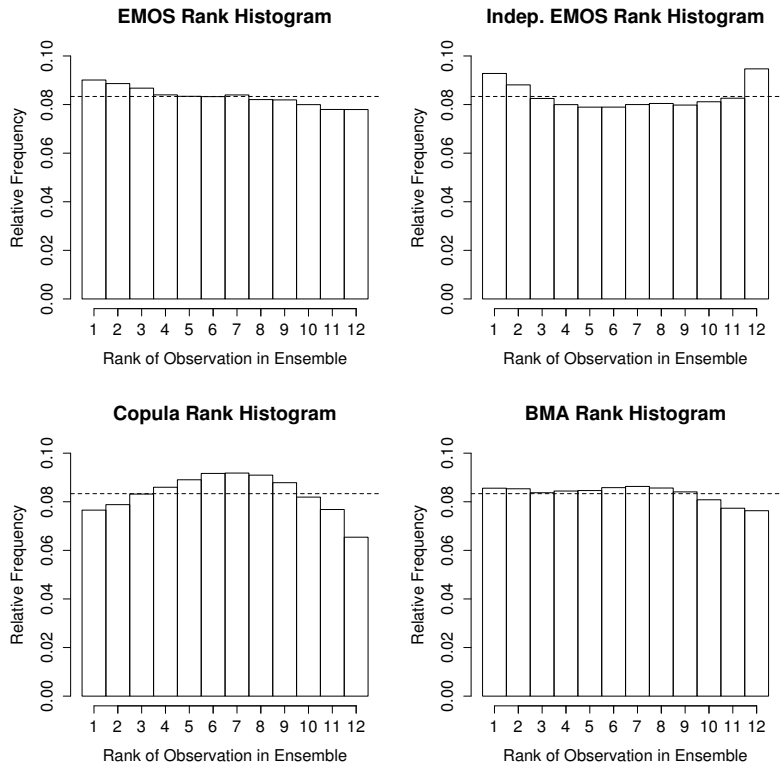,height=10cm, angle=0}
\caption{Multivariate rank histograms for EMOS ({\em upper left}),
  independent EMOS ({\em upper right}), Gaussian copula ({\em lower
    left}) and BMA ({\em lower right})
  post-processed ALADIN-HUNEPS forecasts of instantaneous wind speed
  and temperature.}
\label{fig:fig4}
\end{center}
\end{figure}

The effects of statistical calibration of ensemble forecasts are
quantified by the multivariate scores given in Table
\ref{tab:tab2}. Compared to the raw ensemble all four post-processing
methods result in significantly lower energy scores and reliability
indices (compare Figures \ref{fig:fig3} and \ref{fig:fig4}) and higher
$\DS$ values. Again, the loss in determinant
sharpness is an effect of the underdispersive nature of the
ensemble. However, here the increase in
$\DS$ is around $60\,\%$, whereas for the UWME the raw ensemble is
almost three times sharper than the various predictive PDFs. This
again indicates the better calibration of the ALADIN-HUNEPS ensemble
which is fully consistent with Figures \ref{fig:fig1} and \ref{fig:fig3} and
the corresponding reliability indices given in Sections
\ref{subsub4.1.1} and \ref{subsub4.2.1}, respectively. Further, the
ensemble median and mean vectors produce slightly larger Euclidean errors than
the corresponding post-processed point forecasts. Moreover, the empirical
correlations of the components of the ensemble median and mean are
almost the double of the nominal correlation $-0.033$ of observations,
whereas the correlations of wind speed and temperature components of
the BMA and EMOS point forecasts are close to this value. Finally,
both the ensemble median/mean and their calibrated counterparts
exhibit almost the same forecast error correlations.

From the competing post-processing methods the Gaussian copula
approach results in the lowest energy score and Euclidean errors, however, the
differences compared to the corresponding scores of the BMA and EMOS models
(especially in the $\EE$ values) are rather small. Reliability indices
show far larger variability and the highest scores belong to the
copula model and to the independent EMOS approach. The \ $\Delta$ \
values in Table \ref{tab:tab2} are in accordance with the rank
histograms in Figure \ref{fig:fig4}: the rank histogram of the copula
method is strongly hump-shaped indicating over-dispersion \citep[see,
e.g.,][]{gsghj}, whereas the histogram of the independent EMOS
approach exhibits some under-dispersion. For the ALADIN-HUNEPS
ensemble the bivariate BMA model has the best overall performance
closely followed by the bivariate EMOS method, however, similar to the
case of the UWME the computational costs might also effect the model choice.

\subsection{Computational aspects}
\label{subs:subs4.3}

For all EMOS methods which have been developed so far the most
time-consuming and problematic part of ensemble post-processing is the
numerical optimization used in parameter estimation. In case of
bivariate EMOS calibration of the ALADIN-HUNEPS ensemble only the
robust Nelder-Mead algorithm occurs to be
reliable, as one has to estimate 18 free parameters with the help of
400 forecast cases of the training data. For the UWME the
data/parameter ratio is much better, as 42 free parameters have to be
estimated using on average 3354 forecast cases. For this data set the
reported Nelder-Mead and the faster BFGS algorithm give almost the same
results.

In case of the BMA calibration the bottleneck with respect to the
computation costs is the EM algorithm applied for ML estimation of the
parameters. The bivariate BMA model of \citet{bm} makes use of a
modification of the truncated data EM algorithm for Gaussian mixture
models \citep{ls12} which operates with closed formulae and there is no
need of numerical optimization. However, due to the large number of
free parameters (UWME: 59; ALADIN-HUNEPS: 17) it requires quite a lot
of iterations resulting in long computation times.

The Gaussian copula method starts with very fast univariate EMOS
calibration, however, it utilizes an additional data set for
estimating the correlation matrix of the Gaussian copula and
additional post-processing of the
univariate predictive PDFs. Hence, in terms of computational efficiency
this method is not comparable with the bivariate approaches and it is
excluded from our analysis.

\begin{figure}[t!]
\centering
\centerline{
\hbox to 13.5 truecm{
\epsfig{file=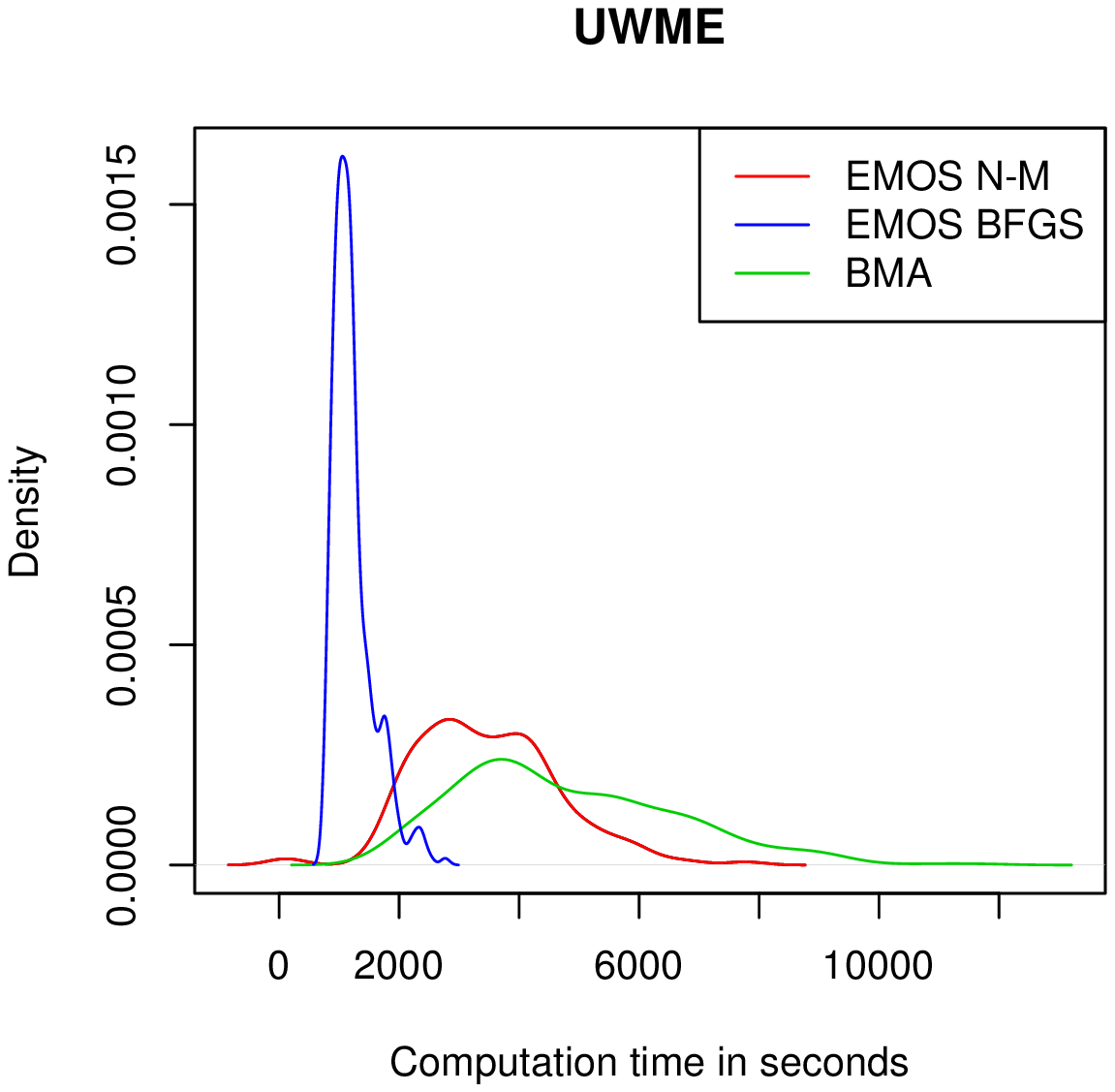,height=6.5cm, angle=0} \hfill
\epsfig{file=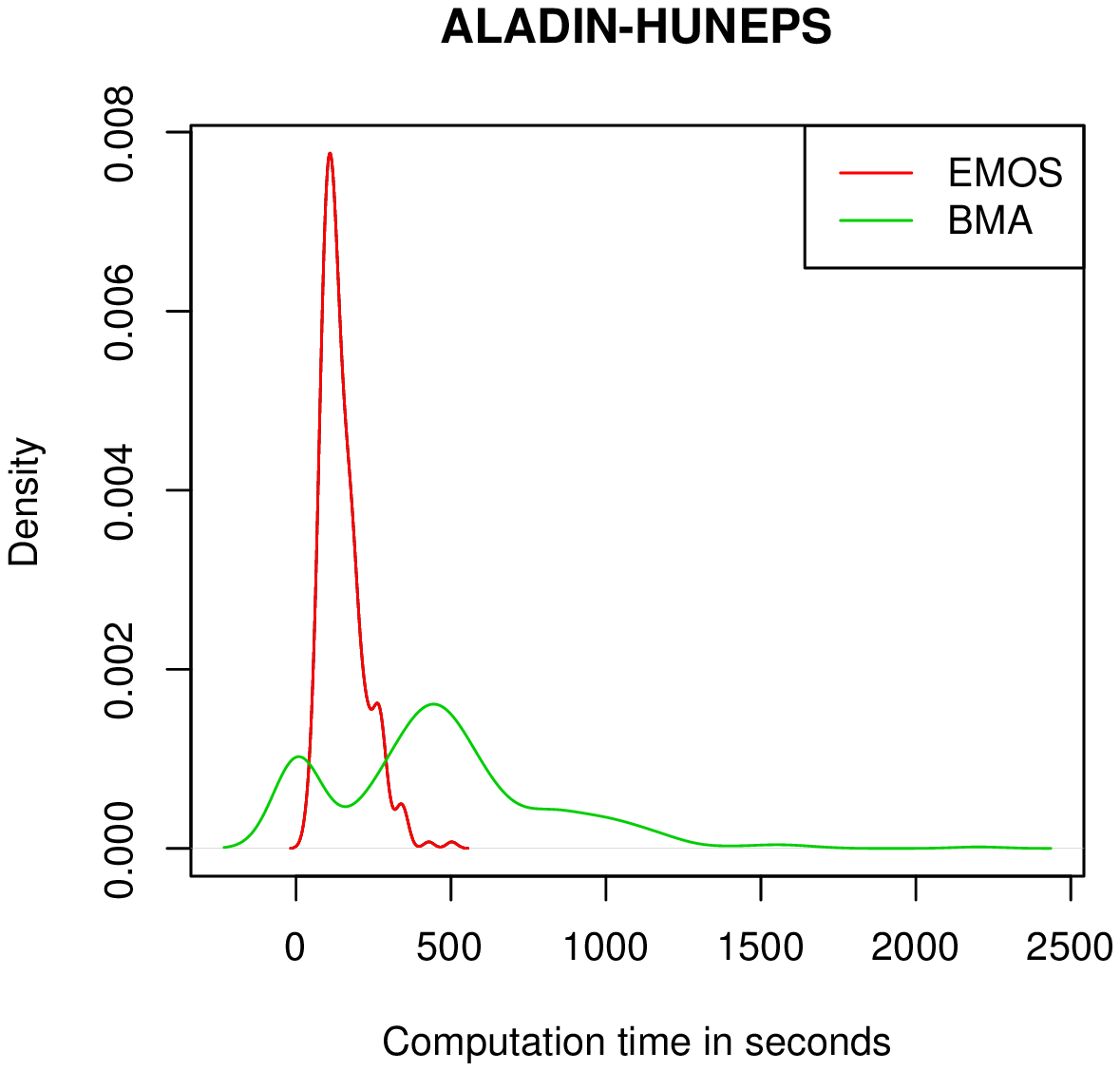,height=6.5cm, angle=0}}}
\centerline{\hbox to 7.5cm{\small \quad(a) \hfill  (b)
    \hspace{-0.75cm}}}
\caption{Densities of computation times for the bivariate BMA and EMOS
  models.  a) UWME  for the calendar year 2008; b) ALADIN-HUNEPS
  ensemble for the period May 12, 2012 -- March 31, 2013.}
\label{fig:fig5}
\end{figure}
Figures \ref{fig:fig5}a and  \ref{fig:fig5}b show the kernel density 
estimates of the
distribution of computation times over the days in the verification
period for bivariate BMA and EMOS models (implemented in {\tt R}) for
the UWME and ALADIN-HUNEPS ensemble, respectively, calculated on a
portable computer under a 64 bit Fedora 20 operating
system (Intel Quad Core i7-4700MQ CPU (2.40GHz $\times$ 4), 20 Gb
RAM). We remark that in Figure \ref{fig:fig5}a
the density of computation times of the EMOS model with BFGS
optimization is also plotted.
The densities displayed in Figure \ref{fig:fig5} clearly show
that in terms of computation time the EMOS model outperforms the BMA
approach. The same conclusion can be derived from Table \ref{tab:tab3}
where the median, mean and standard deviation of the computation times are
given. However, one should also remark that these computation times
are still too long for an operational use.

\begin{table}[t!]
\begin{center}{
\begin{tabular}{lccccc}
&\multicolumn{3}{c}{UWME}&\multicolumn{2}{c}{
 ALADIN-HUNEPS}\\\cline{2-6}
Model&\multicolumn{2}{c}{EMOS}&BMA&EMOS&BMA \\ \cline{2-3} \cline{5-5}
&Nelder-Mead&BFGS& &Nelder-Mead& \\
\hline
median&3349.443 &1140.702 &4419.288 &131.609 &436.873 \\
mean&3475.681 &1228.142 &4801.247 &150.008 &459.560 \\
std. dev.&1177.651 &343.633 &1823.083 &69.678 &345.187 \\
\end{tabular}
\caption{Median, mean and standard deviation of the computation times
  in seconds allocated to the parameter estimation for individual days
  in the verification period
  (UWME: calendar year 2008, 24\,302 forecast cases;
  ALADIN-HUNEPS ensemble: May 12, 2012 -- March 31, 2013, 3\,180
  forecast cases).} \label{tab:tab3}
}
\end{center}
\end{table}

\section{Conclusions}
  \label{sec:sec5}

We introduce a new EMOS model for joint calibration of ensemble forecasts of
wind speed and temperature providing a predictive PDF which follows a
bivariate normal distribution truncated from below at zero in its first
coordinate. The model is tested on wind speed and temperature forecasts of
the eight-member University of Washington mesoscale ensemble and of
the eleven-member ALADIN-HUNEPS ensemble of the Hungarian
Meteorological Service. These ensemble prediction systems differ both
in the weather quantities being forecasted and in the generation of the ensemble
members. 

Using appropriate verification measures (energy score,
reliability index and determinant sharpness of probabilistic and
Euclidean errors, correlations, as well as correlations of errors of
median/mean forecasts) the predictive
performance of the bivariate EMOS model is compared to the forecast
skills of the independent EMOS calibration of wind speed and
temperature, the Gaussian copula method of \citet{mlt} based on
univariate EMOS models, the bivariate BMA model suggested by
\citet{bm} and the raw ensemble vectors as well.

From the results of the presented case studies one can conclude that
compared to the raw ensemble post-processing always improves the calibration of
probabilistic and accuracy of point forecasts. Further, in terms of
predictive performance the bivariate
EMOS model is able to keep up with the other two bivariate
methods. Concerning the  
computational costs it outperforms the bivariate BMA method, whereas
compared to the Gaussian copula approach it does not require an
additional data set for estimating the correlations.

\bigskip
\noindent
{\bf Acknowledgments.} \  \
Essential part of this work was made during the visit of S\'andor Baran
at the Heidelberg Institute for Theoretical Studies. The research stay in 
Heidelberg was funded by the DAAD program ``Research Stays for University 
Academics and Scientists, 2015''. S\'andor Baran was also supported by
the J\'anos Bolyai Research Scholarship of the Hungarian Academy of Sciences.
The authors are indebted to Tilmann
Gneiting for his useful suggestions and
remarks, to the University of Washington MURI group for providing the
UWME data, to Mih\'aly Sz\H ucs from the HMS for the
ALADIN-HUNEPS data and to Thordis Thorarinsdottir and Alex Lenkoski for their
help with the R codes for the copula method.


\begin{thebibliography}{99}
\bibitem[Bao  {\em et al.\/}, 2010]{bgrgg} Bao, L., Gneiting, T.,
  Raftery, A. E., Grimit, E. P. and Guttorp, P. (2010) Bias
  correction and Bayesian model averaging for ensemble forecasts of
  surface wind direction. {\em Mon. Wea. Rev.\/} {\bf 138},
  1811--1821.

\bibitem[Baran, 2014]{bar} Baran, S. (2014) Probabilistic wind speed
  forecasting using Bayesian
   model averaging with truncated normal components. {\em
     Comput. Stat. Data. Anal.\/} {\bf 75}, 227--238.

\bibitem[Baran {\em et al.\/}, 2013]{bhn1}  Baran, S., Hor\'anyi, A. and
  Nemoda, D. (2013) Statistical
  post-processing of probabilistic wind speed
  forecasting in Hungary. {\em Meteorol. Z.\/}  {\bf 22},
  273--282.

\bibitem[Baran {\em et al.\/}, 2014]{bhn2}  Baran, S., Hor\'anyi, A. and
  Nemoda, D. (2014) Probabilistic
   temperature forecasting with statistical calibration in Hungary.
   {\em  Meteorol. Atmos. Phys.\/} {\bf 124}, 129--142.

 \bibitem[Baran and Lerch, 2015]{bl} Baran, S., Lerch, S. (2015)
   Log-normal distribution based EMOS models for probabilistic wind
   speed forecasting. {\em Q. J. R. Meteorol. Soc.\/}, doi:10.1002/qj.2521.

 \bibitem[Baran and M\"oller, 2015]{bm} Baran, S., M\"oller, A. (2015) Joint
   probabilistic
  forecasting of wind speed and temperature using Bayesian model
  averaging. {\em Environmetrics\/} {\bf 26}, 120--132.

\bibitem[Bouall\`egue {\em et al.\/}, 2013]{btg} Bouall\`egue, B. Z.,
  Theis, S. and Gebhardt, C. (2013) Enhancing COSMO-DE ensemble forecasts
  by inexpensive techniques.  {\em Meteorol. Z.\/} {\bf 22}, 49--59.

\bibitem[Buizza  {\em et al.\/}, 1993]{btmp} Buizza, R., Tribbia, J.,
  Molteni, F. and  Palmer, T. (1993) Computation of optimal unstable
  structures for a numerical weather prediction system.  {\em Tellus
    A} {\bf 45}, 388--407.

\bibitem[Buizza  {\em et al.\/}, 2005]{bhtp} Buizza, R., Houtekamer,
  P. L., Toth, Z., Pellerin, G., Wei, M. and Zhu, Y. (2005) A
  comparison of the ECMWF, MSC, and NCEP global ensemble prediction
  systems.  {\em Mon. Wea. Rev.\/} {\bf 133}, 1076--1097.

\bibitem[Delle Monache {\em et al.\/}, 2006]{delle} Delle Monache, L.,
  Hacker, J. P., Zhou, Y., Deng, X. and Stull, R. B. (2006)
Probabilistic aspects of meteorological and ozone regional ensemble
forecasts. {\em J. Geophys. Res.\/} {\bf 111} D24307.

\bibitem[Dennis and Schnabel, 1983]{ds83} Dennis, J. and Schnabel,
  R. (1983) {\em Numerical Methods for Unconstrained Optimization and
Nonlinear Equations.\/} Prentice Hall, New Jersey.

\bibitem[Descamps {\em et al.\/}, 2014]{dljbac}  Descamps, L., Labadie, C.,
  Joly, A., Bazile, E., Arbogast, P. and C\'ebron, P. (2014). PEARP, the
  M\'et\'eo-France short-range ensemble prediction system. {\em
    Q. J. R. Meteorol. Soc.\/}, doi:10.1002/qj.2469.

\bibitem[Eckel and Mass, 2005]{em05} Eckel, F. A. and Mass,
  C. F. (2005) Effective mesoscale, short-range ensemble forecasting. {\em
    Wea. Forecasting\/} {\bf 20}, 328--350.

\bibitem[ECMWF Directorate, 2012]{ecmwf} ECMWF Directorate
  (2012) Describing ECMWF's forecasts and forecasting system. {\em
    ECMWF Newsletter\/} {\bf 133}, 11--13.

\bibitem[Fraley {\em et al.\/}, 2010]{frg} Fraley, C., Raftery,
  A. E. and Gneiting, T. (2010) Calibrating multimodel forecast
  ensembles with exchangeable and missing members using Bayesian model
  averaging. {\em Mon. Wea. Rev.\/} {\bf 138}, 190--202.

\bibitem[Fraley {\em et al.\/}, 2011]{frgsb} Fraley, C., Raftery,
  A. E., Gneiting, T., Sloughter, J. M. and Berrocal, V. J. (2011)
  Probabilistic weather forecasting in R.  {\em The R Journal\/} {\bf
    3}, 55--63.

\bibitem[Fritz {\em et al.\/}, 2012]{ffc} Fritz, H., Filzmoser, P. and
  Croux, C. (2012) A comparison of algorithms for the multivariate
  $L_1$-median. {\em Comput. Stat.\/} {\bf 27}, 393--410.

\bibitem[Gneiting, 2014]{gneiting14} Gneiting, T. (2014) Calibration
  of medium-range weather forecasts. {\em ECMWF Technical
    Memorandum\/} No. 719. Available at:
  old.ecmwf.int/publications/library/ecpublications/
  \_{}pdf/tm/701-800/tm719.pdf. Accessed 27 July 2015.

\bibitem[Gneiting and Raftery, 2005]{gr} Gneiting, T. and Raftery,
  A. E. (2005) Weather forecasting with ensemble methods. {\em
    Science\/} {\bf 310}, 248--249.

\bibitem[Gneiting and Raftery, 2007]{grjasa} Gneiting, T. and Raftery,
  A. E. (2007) Strictly proper scoring rules, prediction and
  estimation. {\em J. Amer. Statist. Assoc.\/} {\bf 102}, 359--378.

\bibitem[Gneiting {\em et al.\/}, 2005]{grwg} Gneiting, T.,
  Raftery, A. E., Westveld, A. H. and Goldman, T. (2005) Calibrated
  probabilistic forecasting using ensemble model output statistics and
  minimum CRPS estimation. {\em Mon. Wea. Rev.\/} {\bf 133},
  1098--1118.

\bibitem[Gneiting {\em et al.\/}, 2008]{gsghj} Gneiting, T.,
  Stanberry, L. I., Grimit, E. P., Held, L. and Johnson, N. A. (2008)
  Assessing probabilistic forecasts of multivariate quantities, with
  applications to ensemble predictions of surface winds (with
  discussion and rejoinder). {\em Test\/} {\bf 17}, 211--264.

\bibitem[Grell {\em et al.\/}, 1995]{grell} Grell, G. A., Dudhia,
  J. and Stauffer, D. R. (1995) A description of the fifth-generation
  Penn state/NCAR mesoscale model (MM5). {\em Technical Note\/}
  NCAR/TN-398+STR. National Center for Atmospheric Research,
  Boulder. Available at:
  http://nldr.library.ucar.edu/repository/assets/technotes/TECH-NOTE-000-000-000-214.pdf. Accessed 27 July 2015.
 

\bibitem[Hor\'anyi {\em et al.\/}, 2006]{hkkr} Hor\'anyi, A,
  Kert\'esz, S., Kullmann, L. and Radn\'oti, G. (2006) The
  ARPEGE/ALADIN mesoscale numerical modelling system and its
  application at the Hungarian Meteorological Service.  {\em Id\H
    oj\'ar\'as\/} {\bf 110}, 203--227.

\bibitem[Hor\'anyi {\em et al.\/}, 2011]{horanyi} Hor\'anyi, A., Mile, M.,
  Sz\H ucs, M. (2011) Latest developments around the ALADIN operational
  short-range
  ensemble prediction system in Hungary. {\em Tellus A} {\bf 63},
  642--651.

\bibitem[Lee and Scott, 2012]{ls12} Lee, G. and Scott, C. (2012) EM
  algorithms for multivariate Gaussian mixture models with truncated
  and censored data. {\em Comput. Statist. Data Anal.\/}  {\bf 56},
  2816--2829.

\bibitem[Leith, 1974]{leith} Leith, C. E. (1974) Theoretical
  skill of Monte-Carlo forecasts.  {\em Mon. Wea. Rev.\/} {\bf 102},
  409--418.

\bibitem[Lerch and Thorarinsdottir, 2013]{lt} Lerch, S. and
  Thorarinsdottir, T. L. (2013) Comparison of non-homogeneous
  regression models for probabilistic wind speed forecasting. {\em
    Tellus A\/} {\bf 65}, 21206.

\bibitem[Leutbecher and Palmer, 2008]{lp} Leutbecher, M. and Palmer,
  T. N. (2008) Ensemble forecasting. {\em J. Comp. Phys.\/}  {\bf 227},
  3515--3539.

\bibitem[Milasevic and Ducharme, 1987]{md} Milasevic, P. and Ducharme,
  G. R. (1987) Uniqueness of the spatial median. {\em Ann. Statist.\/}
  {\bf 15}, 1332--1333.

\bibitem[M\"oller {\em et al.\/}, 2013]{mlt} M\"oller, A., Lenkoski,
  A. and Thorarinsdottir, T. L. (2013) Multivariate
  probabilistic forecasting using ensemble Bayesian model averaging
  and copulas. {\em Q. J. R. Meteorol. Soc.\/} {\bf 139}, 982--991.

\bibitem[National Weather Service, 1998]{asos} National Weather
  Service (1998) {\em Automated Surface Observing System
(ASOS) User’s Guide.\/} Available at:
http://www.weather.gov/asos/aum-toc.pdf. Accessed 27 July 2015.

\bibitem[Nelder and Mead, 1965]{nm} Nelder, J. A. and Mead, R. (1965)
  A simplex algorithm for function minimization. {\em Comput. J.\/}
  {\bf 7}, 308--313.

\bibitem[Pinson, 2012]{pinson} Pinson, P. (2012) Adaptive calibration
  of $(u,v)$-wind ensemble forecasts. {\em Q. J. R. Meteorol. Soc.\/}
  {\bf 138}, 1273--1284.

\bibitem[Press {\em et al.\/}, 2007]{press} Press, W. H., Teukolsky,
  S. A., Vetterling, W. T. and Flannery, B. T. (2007) {\em Numerical
    Recipes 3rd Edition: The Art of Scientific Computing.\/} Cambridge
  University Press, Cambridge.

\bibitem[Raftery {\em et al.\/}, 2005]{rgbp} Raftery, A. E., Gneiting, T.,
  Balabdaoui, F. and Polakowski, M. (2005) Using Bayesian model
  averaging to calibrate forecast ensembles. {\em Mon. Wea. Rev.\/}
  {\bf 133}, 1155--1174.

\bibitem[Rosenbaum, 1961]{rose} Rosenbaum, S. (1961) Moments of a
  truncated bivariate normal distribution. {\em J. Roy.
    Statist. Soc. Ser. B\/} {\bf 23}, 405--408.

\bibitem[Schefzik {\em et al.\/}, 2013]{stg13} Schefzik, R.,
  Thorarinsdottir, T. L. and Gneiting, T. (2013) Uncertainty
  quantification in complex simulation models using ensemble copula
  coupling. {\em Statist. Sci.\/} {\bf 28}, 616--640.

\bibitem[Scheuerer, 2014]{sch} Scheuerer, M. (2014) Probabilistic
  quantitative precipitation forecasting using ensemble model output
  statistics. {\em Q. J. R. Meteorol. Soc.\/} {\bf 149}, 1086--1096.

\bibitem[Schuhen {\em et al.\/}, 2012]{stg} Schuhen, N.,
  Thorarinsdottir, T. L. and Gneiting, T. (2012) Ensemble model output
  statistics for wind vectors. {\em Mon. Wea. Rev.\/}
  {\bf 140}, 3204--3219.

\bibitem[Sloughter {\em et al.\/}, 2010]{sgr10} Sloughter,
  J. M., Gneiting, T. and Raftery, A. E. (2010)  Probabilistic wind
  speed forecasting using ensembles and Bayesian model averaging. {\em
    J. Amer. Stat. Assoc.\/} {\bf 105}, 25--37.

\bibitem[Sloughter {\em et al.\/}, 2013]{sgr13} Sloughter,
  J. M., Gneiting, T. and Raftery, A. E. (2013)  Probabilistic wind
  vector forecasting using ensembles and Bayesian model averaging. {\em
     Mon. Wea. Rev.\/} {\bf 141}, 2107--2119.

\bibitem[Thorarinsdottir and Gneiting, 2010]{tg}  Thorarinsdottir,
  T. L. and Gneiting, T. (2010) Probabilistic forecasts of wind speed:
  ensemble model output statistics by using heteroscedastic censored
  regression. {\em J. Roy. Statist. Soc. Ser. A\/} {\bf 173},
  371--388.

\bibitem[Toth and Kalnay, 1997]{tk} Toth, Z. and Kalnay, E. (1997)
  Ensemble forecasting at NCEP and the breeding method.  {\em
    Mon. Wea. Rev.\/} {\bf 125},  3297--3319.

\bibitem[Vardi and Zhang, 2000]{vz} Vardi, Y. and Zhang, C. H. (2000)
  The multivariate $L_1$-median and associated data depth. {\em
    Proc. Natl. Acad. Sci. USA\/} {\bf 97}, 1423--1426.

\bibitem[Wilks, 2011]{wilks} Wilks, D. S. (2011) {\em Statistical
    Methods in the Atmospheric Sciences.\/} 3rd ed., Elsevier,
  Amsterdam.

\bibitem[Williams {\em et al.\/}, 2014]{wfk} Williams, R. M., Ferro,
  C. A. T. and Kwasniok, F. (2014) A comparison of ensemble
  post-processing methods for extreme events. {\em
    Q. J. R. Meteorol. Soc.\/} {\bf 140},  1112--1120.
\end{thebibliography}
\end{document}